\documentclass[preprint,12pt]{elsarticle}
\journal{Chinese Journal of Physics}
\usepackage{graphicx}

\usepackage{lineno,hyperref}
\modulolinenumbers[5]
\usepackage{array}
\usepackage{microtype}
\usepackage{float}
\usepackage{amsfonts}
\usepackage{mathrsfs}
\usepackage{amsmath}
\usepackage{epstopdf}
\hypersetup{colorlinks,citecolor=blue}
\begin{document}

\begin{frontmatter}

\title{Analytical study of ion-acoustic solitary waves in a magnetized plasma with degenerate electrons}

\author[addr1]{Moumita Indra\footnote{Orcid ID: 0000-0002-7900-7947}}\ead{moumita.indra93@gmail.com}
\author[addr2]{K. K. Ghosh}\ead{kkghosh1954@gmail.com}
\author[addr3]{Saibal Ray\footnote{Orcid ID: 0000-0002-5909-0544}}\ead{saibal.ray@gla.ac.in}

\address[addr1]{School of Basic Science, Swami Vivekananda University, Kanthalia, Barrackpore, West Bengal, India}
\address[addr2]{Department of Basic Science \& Humanities, Abacus Institute of Engineering and Management, Magra, Chinchura, West Bengal, India}
\address[addr3]{Centre for Cosmology, Astrophysics and Space Science (CCASS), GLA University, Mathura 281406, Uttar Pradesh, India}

\date{Received~~2022 September 22; accepted~~2022~~month day}

\begin{abstract}
The propagation of fully nonlinear ion acoustic solitary waves (IASW) in a magneto-plasma with degenerate electrons investigated by Abdelsalam et al.~[\cite{AS} {\color{blue}Physics Letters A 372 (2008) 4923}]. Based on their work in the present work, a rigorous and general analytical study is presented. This confirms their implied assumption that (i) only hump and no cavity is possible and (ii) for humps, the algebraic equation for the maximum density $N$ obtained by them determines it uniquely (naturally assumes $N > 1$). Here we confirm analytically their assertion that $N$ decreases with $l_x$ (the direction cosine of the wave vector $k$ along the $x$-axis) and $N$ increases with the increase of the Mach number ($M$).

Keywords: Waves; Plasmas; MHD; Hydrodynamics; Mach number
\end{abstract}

\end{frontmatter}

\section{Introduction}
Recently, the physics of an electron-positron-ion (EPI) plasma~\cite{e-p-i1992a,e-p-i1992b,e-p-i1994a,e-p-i1994b,epia,epib} has received considerable attention, mainly due to its importance in many systems in laboratory plasma as well as astrophysical arena. EPI plasma exists in places such as active galactic nuclei (AGNs)~\cite{Gallac_nuclei,Piotrovich2020}, pulsar magneto-spheres~\cite{magnetospherea,magnetosphereb} and in many dense astronomical environments, namely, neutron stars and white dwarfs~\cite{NS_WDa,NS_WDb} which is supposed to play a key role in understanding the origin and evolution of our entire universe~\cite{universe}. This kind of plasma may also be practically produced in laboratories~\cite{laba,labb,labc,labd}. 

Electrons and positrons are assumed relativistic and degenerate, following the Fermi–Dirac statistics, whereas the warm ions are described by a set of classical fluid equations with an individual charge of $Z_i e$, ($Z_i$ denotes the ion-charge state, while $e$ is the electron charge), subject to the influence of the electrostatic potential $\phi$. Quantum hydrodynamics ~\cite{QHDa,QHDb,QHDc,QHDd}, which describes quantum systems within a hydrodynamic framework, was first proposed by Madelung~\cite{Madelung} and Bohm~\cite{Bohm}. Although such a description is formally accurate for a single particle, Manfredi and Haas~\cite{Haas2001} later expanded the idea to many-particle systems and it gained significant favour in the areas of the quantum plasma community. 

Considering the one-dimensional QHD model in the limit of the small mass ratio of the charge carriers, Hass et al.~\cite{Haas} were the first to study the ion-acoustic waves in unmagnetized quantum plasma. This model has been used in various investigations by several authors~\cite{QHDa,QHDb,QHDc,QHDd} where generally a linear dispersion relation is derived in the linear approximation. Thus, ion-acoustic waves (IAW), a fundamental mode in plasma environments, have been a subject of extensive research over several decades. Khan and Haque~\cite{Khan&Haque} showed that in the small (linear) limit of quantum diffraction parameter $H$ (ratio of the plasmon energy to the Fermi energy), the system behaves as the classical IAW whereas in the non-linear regime the system behaves differently. One of the most interesting non-linear features of IAW is the existence of ion-acoustic solitary waves (IASW)~\cite{IASWa,IASWb}. 

In the weakly nonlinear limit, the quantum plasma is shown to support waves described by a deformed Korteweg–de Vries (KdV) equation which depends in a non-trivial way on the quantum parameter $H$. However, in the fully non-linear regime, the system exhibits travelling waves which show a periodic pattern. Hence there are two main approaches used to investigate IASWs, viz., the reductive perturbation technique (KdV method)~\cite{KdV} and the pseudo-potential technique for large-amplitude solitary waves (Sagdeev method)~\cite{Sagdeev}. The theory of solitons in magneto-plasma was greatly improved by an intriguing work by Abdelsalam et al.~\cite{AS} on completely non-linear IASW travelling obliquely to an external magnetic field in a collision-less dense Thomas-Fermi magneto-plasma with degenerate electrons. 

The degenerate electrons in the above scenario may be described using the Thomas-Fermi approximation~\cite{ThomasFermia,ThomasFermib} whereas the ion component can be thought of as a classical gas. They have obtained an energy balance-like equation involving the Sagdeev potential as follows:
\begin{equation}
\frac{1}{2}\left(\frac{dn}{d\eta}\right)^2 + V(n) = 0,
\label{1}
\end{equation}
where Sagdeev-like pseudo-potential $V(n)$ is given by
\begin{align}
V(n)=\frac{9n^6}{2(5an^{8/3}-3)^2}[\frac{5an^{8/3}-2an^{5/3-3a+1}}{M^2} \nonumber\\ +\frac{1-2n}{M^2n^2} + cM^2(an^{10/3}-2an^{5/3}-\frac{2}{9}+a+5)],
\label{2}
\end{align}
where
\begin{equation}
a = \frac{3}{5M^2},
\label{3}
\end{equation}

\begin{equation}
c = \frac{3{l_x}^2}{5M^2},
\label{4}
\end{equation}

\begin{align*}
\eta=l_x x +l_y y -Mt,~{l_x}^2 + {l_y}^2=1~\mbox{and}~n=n(\eta).
\end{align*}

Here $M$ is the Mach number, $l_x$ and $l_y$ are the direction cosines of the wave vector $k$ along the $x$ and $y$ axes respectively, $n(=\frac{n_e}{n_o})$ where $n_e$ is the electron density and $n_o$ is the unperturbed electron density with $n_i$ as the ion density.

As indicated by Abdelsalam et al.~\cite{AS}, the existence of IASW's for which
\begin{align}
 1\le n\le N~\mbox{and} \; \frac{dn}{d\eta} = 0 \; \mbox{at} \; n=1,N,
 \label{5}
 \end{align} 
requires the following equations and inequality:
\begin{equation}
V(n)|_{n=1}=0, 
\label{6}
\end{equation}

\begin{equation}
\frac{dV}{dn}|_{n=1}=0,
\label{7}
\end{equation}

\begin{equation}
\frac{d^2V}{dn^2}|_{n=1} < 0,  
\label{8}
\end{equation}

\begin{equation}
\mbox{and}\; V(n)|_{n=N}=0.
\label{9}
\end{equation}

Abdelsalam et al.~\cite{AS} noted that Eqs. (\ref{6}) and (\ref{7}) are automatically satisfied by Eq. (\ref{2}) while the inequality (\ref{8}) is satisfied if and only if
\begin{align*}
l_x <M<1,~\mbox{i.e.,} \;\; c<0.6<a,
\end{align*}
in view of Eqs. (\ref{3}) and (\ref{4}).

For the nonlinear dispersion relation, Eq. (\ref{9}), they have numerically solved it for several specific values of $l_x$ ($l_x= 0.66, 0.68, 0.7$) and on that basis argued that if Eq. (\ref{9}) can be rewritten as
\begin{equation}
N=N(l_x,M),
\label{10}
\end{equation}
where the maximum density $N$ is a decreasing function of $l_x$, i.e., $\frac{\partial N}{\partial l_x} < 0$ and $N$ is an increasing function of $M$, i.e., $\frac{\partial N}{\partial M} > 0$.

However, in the present work our motivation is to solve the problem of Abdelsalam et al.~\cite{AS} with an  analytical methodology under a more general treatment. For this we have considered a different format and have shown that some of their outcomes can be retrieved with a convincing way and can be demonstrated valid in the physical realm.

\section{An analytical methodology} 

Putting $n=x^3$ the equations (\ref{1}) and (\ref{2}) can be rewritten as
\begin{equation}
\left(\frac{dx}{d\eta}\right)^2 + \frac{x(1-x)f(x,l_x, M)}{(5ax^8-3)^2} =0,
\label{11}
\end{equation}
where 
\begin{align}
f(x,l_x,M) = (1+x+x^2)^2 \nonumber \\ +\frac{9{l_x}^2 x^6 (1+x+x^2+x^3+x^4)^2}{25M^4} \nonumber \\
-\frac{3x^6(3+6x+4x^2+2x^3)}{5M^2} \nonumber \\ - \frac{3{l_x}^2x^3 (2+4x+6x^2+3x^3)}{5M^2}.
\label{12}
\end{align}

Equations (\ref{5}) and (\ref{9}) are now rewritten as 
\begin{align}
\frac{dx}{d\eta}=0  \;\mbox{at}\; x=1, N^{1/3},\\
f(N^{1/3}, l_x, M) = 0.
\label{14}
\end{align}

The above Eqs. (\ref{3}) and (\ref{4}) and the inequality (\ref{8}) remain unchanged except Eq. (\ref{9}) which is to be replaced by Eq. (\ref{14}). In other words, the question now is whether Eq. (\ref{14}) can determine $N$ (or $x$) uniquely. To answer this one needs the following observations on $f(x,l_x,M)$.

\section{Observations on $f(x,l_x,M)$}

\subsection{Observation 1:}
(i) $f(0,l_x,M) = 1$,  \\
(ii) $f(1) = (3-5a)(3-5c)<0$,  \\
(iii) $f(\infty)>0$. \\
\\
\textbf{Proof:} Trivial.

\subsection{Observation 2:}
For given $a$ and $c$ there exist a unique pair of $(\alpha, \beta)$ such that
\begin{align*}
f(x,l_x,M) > 0, \;\mbox{for} \; 0<x<\beta, \\
f(\beta,l_x,M) = 0 , \\
f(x,l_x,M) < 0 , \;\mbox{for} \; \beta < x < \alpha, \\
f(\alpha,l_x,M) = 0 , \\
\mbox{and} \; f(x,l_x,M) > 0 \;\mbox{for} \; x>\alpha,
\end{align*}
where $0<\beta<1<\alpha$. \\

\textbf{Corollary:}
\begin{equation*}
\frac{\partial f(x, l_x, M)}{\partial x} > 0 \; \mbox{at} \; x=\alpha. \nonumber
\end{equation*}

\textbf{Proof:} Trivial.

\subsection{Observation 3:}
\begin{equation*}
f(x, l_x, M) < 0 \; \mbox{at} \; x=(\frac{3}{5a})^{1/8}.  \nonumber
\end{equation*}

\textbf{Proof:} See Appendix. \\
\\
\textbf{Corollary:}
\begin{equation*}
\beta < (\frac{3}{5a})^{1/8} <1.
\end{equation*}\\

\textbf{Proof:} Trivial from observation 2. \\

\subsection{Observation 4:}
\begin{equation*}
\frac{\partial f(x, l_x, M)}{\partial l_x} > 0 \; \mbox{for} \; x>1.  \nonumber
\end{equation*}\\
\\
\textbf{Proof:} See Appendix.

\subsection{Observation 5:}
For any $\alpha>1$ there exists $l_x$ and $M$ that satisfy $f(\alpha, l_x, M)=0$ and also satisfy the inequality (\ref{9}).\\
\\
\textbf{Proof:} See Appendix. \\

With these observations one can uniquely determine $x$ (or $N$) ($>1$) satisfying Eq. (\ref{12}) and also deals with decreasing/increasing feature of $x$ (or $N$) as well as for increase of $l_x$ or $M$. These are answered as follows. 

\section{Proof of uniqueness of $N$}
From the Observation 1, we note that $f(1) <0$ and $f(\infty)>0$. Owing to the continuity of $f(x)$ there exists one $x$, such that
$f(x_1)=0$ and $x_1 > 1$. If possible, let there exist $x_1$ and $x_2$ such that
\begin{equation}
f(x_1)=f(x_2)=0 \; \mbox{and} \; x_2>x_1>1.
\label{15}
\end{equation}

From Eq. (\ref{15}), applying Rolle's theorem, there exists $x_3$ and $x_4$, such that
\begin{equation}
f^\prime (x_3)=f^\prime(x_4)=0   \; \mbox{and} \; x_2>x_4>x_1>x_3>1. 
\label{16} 
\end{equation}

But $f^\prime(x)$ is a polynomial of degree 13 such that $f^\prime(-\infty)<0$, $f^\prime(0)>0$, $f^\prime(1)<0$ and $f^\prime(\infty)>0$. So we can see that $f^\prime(x)$ vanishes only for $x>1$ which contradicts Eq. (\ref{16}).

Hence there exists a unique $x (>1)$ such that $f(x,l_x, M)=0$, i.e. there exists unique $N(>1)$ satisfying Eq. (\ref{9}).

Now, we have to show analytically that the maximum density $N$ is a decreasing function of $l_x$ and is an increasing function of $M$. Differentiating both sides of Eq. (\ref{12}) with respect to $M$, one gets 
\begin{equation}
 \frac{\partial f}{\partial M} < 0, \;\mbox{for}\; x > 1.  
 \label{17}
 \end{equation} 
 
 For
\begin{eqnarray}
\frac{\partial f}{\partial M} = \frac{6x^3}{25M^5}[-6l_x^2x^3(1+x+x^2+x^3+x^4)^2  +  5x^3 M^2 (3+6x+4x^2+2x^3) \nonumber \\
 + 5l_x^2M^2(2+4x+6x^2+3x^3)] \nonumber \\
< \frac{6x^3}{25M^5} [-6 x^3 (1+x+x^2+x^3+x^4)^2 +  5x^3 (3+6x+4x^2+2x^3) \nonumber \\
+ 5l_x^2 (2+4x+6x^2+3x^3)] \mbox{~~~~(since \;$l_x< M< 1$)} \nonumber \\
= \frac{6x^3}{25M^5} [-18(x^9-x^4)-30(x^7-x^2) - 17(x^8-x^3) - 7(x^8-x) \nonumber \\
-13(x^6-x) - 10(x^{10}-1) -2(x^{10}-x^5) -x^6 -6x^{11}] \nonumber \\
< 0,  \; \;\; \mbox{~for~$x > 1$.~~~~~~~~}
\end{eqnarray}

 From Eqs. (\ref{14}) and (\ref{11}), we obtain 
 \begin{equation}
 \frac{\partial N^{1/3}}{\partial l_x}= -\frac{\frac{\partial f}{\partial l_x}}{\frac{\partial f}{\partial N^{1/3}}} \mbox{\; and \;}  \frac{\partial N^{1/3}}{\partial M}= -\frac{\frac{\partial f}{\partial M}}{\frac{\partial f}{\partial N^{1/3}}}, \nonumber \\
 \end{equation}
 which gives
 \begin{eqnarray*}
   \frac{\frac{\partial N^{1/3}}{\partial l_x}}{\frac{\partial N^{1/3}}{\partial M}} = \frac{\frac{\partial f}{\partial l_x}}{\frac{\partial f}{\partial M}}, 
   \mbox{i.e., \;} \frac{\frac{\partial N}{\partial l_x}}{\frac{\partial N}{\partial M}} = \frac{\frac{\partial f}{\partial l_x}}{\frac{\partial f}{\partial M}}, \\
 \mbox{i.e., \;} \frac{\partial N}{\partial l_x} < 0 \;\;\mbox{for $x> 1$,~~}  \\
 \mbox{~~~~~~(using observation 4 and Eq. (\ref{17})).}
 \end{eqnarray*}
 
 Again from Eq. (\ref{14}), we have
 \begin{eqnarray*}
   \frac{\partial N^{1/3}}{\partial l_x} \; \frac{\partial l_x}{\partial M} \; \frac{\partial M}{\partial N^{1/3}} = -1, \\
   \mbox{and \;} \frac{1}{3}N^{-2/3} \frac{\partial N}{\partial l_x} \frac{5}{3} MC = -\frac{1}{3} N^{-2/3} \frac{\partial N}{\partial M}, \\ \mbox{(by Observation (4))} \\
    \mbox{i.e., \;}  \frac{\partial N}{\partial M} > 0 \; \mbox{for \; $x> 1$ \; \;(since, $\frac{\partial N}{\partial l_x} < 0$).}
 \end{eqnarray*}

\subsection{Proof of Observation 5:}
From the observation 1 one can see that the equation $f(x,l_x,M)=0$ has at least one root between $0$ and $1$ and one root greater then $1$. Also one can note that $f(x,l_x,M)$ regarded as a polynomial in $x$ has two changes of sign and hence by Descarte's rule of sign has at most two positive roots. Hence equation $f(x,l_x,M)=0$ has exactly one root between $0$ and $1$ and exactly one root greater than $1$ which are called $\beta$ and $\alpha$ respectively. The continuity of $f(x,l_x,M)$ ensures that the remaining part of the observation is true.

\section{Conclusion}

In the present work our main motivation was to provide an analytically performed rigorous base of the study of Abdelsalam et al.~\cite{AS} on the propagation of fully non-linear ion-acoustic waves in a collision-less magneto-plasma with degenerate electrons. The outcomes of the investigation are interesting and some explicit features can be exhibited as follows: \\
(1) only hump and no cavity is possible;\\
(2) for humps, (i) the algebraic equation for the maximum density $N$ obtained by them determines it uniquely (under the assumption $N > 1$), (ii) $N$ decreases with $l_x$ (the direction cosine of the wave vector $k$ along the $x$-axis) and (iii) $N$ increases with the increase of the Mach number ($M$). All these results yield simply from the maximum density $N$ which can be uniquely determined by Eq. (\ref{9}) under the constraint  $N>1$.

Another motivation of the present work is related to the astrophysical relevance of an EPI plasma, especially in the cases of AGNs~\cite{Gallac_nuclei,Piotrovich2020}, pulsar magneto-spheres~\cite{magnetospherea,magnetosphereb}, neutron stars and white dwarfs~\cite{NS_WDa,NS_WDb}. A suporting and confirmirmational results of Abdelsalam et al.~\cite{AS} therefore will enhance to understand deeply the structural phenomena occuring in different  astrophysical systems. In this connection we would like to mention the very recent work of Piotrovich et al.~\cite{Piotrovich2020} where they have hypothesized that the AGNs are wormhole mouths rather than supermassive
black holes. Essentially due to bizzare gravitational formation wormholes may emit gamma radiation as a result of a collision of accreting flows inside it. Now the interesting fact is that the radiation has a distinctive spectrum much different from those of jets or accretion discs of AGNs. Hopefully an observation of such radiation via the EPI and hence IASW would serve as evidence of the existence of wormholes.

\section*{Appendix}
\subsection*{Proof of Observation 3}
Let $\left(\frac{3}{5a}\right)^{1/8}=\gamma$
so that $a=\frac{3}{5\gamma^8}$ \\

Then at $x= \gamma$
\begin{eqnarray*}
 f(x,a,c) = (1 + \gamma + \gamma^2)^2 + \frac{3c}{5\gamma^2} (1 + \gamma + \gamma^2 + \gamma^3 + \gamma^4)^2  \nonumber \\
  - \frac{3}{5\gamma^2} (3 + 6\gamma + 4\gamma^2 + 2\gamma^3) - c\gamma^3 (2 + 4\gamma + 6\gamma^2 + 3\gamma^3 \nonumber \\
= \frac{1}{5\gamma^2}[ (5\gamma^2 (1+\gamma + \gamma^2)^2 - 3(3+6\gamma+4\gamma^2+2\gamma^3))  \nonumber \\
 + c (3 (1+\gamma + \gamma^2 + \gamma^3+ \gamma^4)^2 - 5\gamma^5 (2+4\gamma+6\gamma^2+3\gamma^3 ))] \nonumber \\
= \frac{1}{5\gamma^2} [ (-9-18 \gamma -7\gamma^2 + 4 \gamma^3 +15 \gamma^4 +10 \gamma^5 +5 \gamma^6 ) \nonumber \\
+ c (3+6\gamma + 9 \gamma^2 +12 \gamma^3 +15\gamma^4 +2 \gamma^5-11\gamma^6-24\gamma^7-12\gamma^8) ] \nonumber \\
=\frac{1}{5\gamma^2} [(\gamma-1) (9+27\gamma+34\gamma^2+30 \gamma^3+15\gamma^4+5\gamma^5)] \nonumber \\
-c (\gamma-1) (3+9\gamma+18\gamma^2+30\gamma^3+45\gamma^4+47\gamma^5+36\gamma^6+12\gamma^7)    \nonumber \\
= \frac{\gamma-1}{5\gamma^2} [(9+27\gamma+34\gamma^2+30\gamma^3+15\gamma^4+5\gamma^5) \nonumber \\
- c (3+9\gamma+18\gamma^2+30\gamma^3+45\gamma^4+47\gamma^5+36\gamma^6+12\gamma^7)] \nonumber \\
< \frac{\gamma-1}{5\gamma^2}[(9+27\gamma+34\gamma^2+30\gamma^3+15\gamma^4+5\gamma^5) \nonumber \\
-3 (3+9\gamma+18\gamma^2+30\gamma^3+45\gamma^4+47\gamma^5+36\gamma^6+12\gamma^7)] \nonumber \\
\le \frac{\gamma-1}{25\gamma^2}[36+108\gamma+116\gamma^2+60\gamma^3-60\gamma^4-116\gamma^5-108\gamma^6-36\gamma^7]  \nonumber \\
\mbox{~~~~~} <0 \;\mbox{~~~if} \;\gamma<1  \mbox{~~~~~~~~~~~~~~~~~~~~~~~~~~~~~~~~~~~~~~~}\nonumber   
\end{eqnarray*}

\subsection*{Proof of Observation 4}
\begin{eqnarray*}
\frac{\partial f(x,a,c)}{\partial c} = ax^6(1+x+x^2+x^3+x^4)^2-x^3(2+4x+6x^2+3x^3) \nonumber \\
\ge \frac{x^3}{5} [3x^3(1+x+x^2+x^3+x^4)^2 -5(2+4x+6x^2+3x^3)] \;\; \mbox{(since, \;$ a= \frac{3}{5M^2}$)~~~~~~~~~~} \nonumber \\ 
=\frac{x^3}{5} [3x^3 (1+2x+3x^+4x^3+5x^4+4x^5+3x^6+2x^7+x^8) -5(2+4x+6x^2+3x^3)] \nonumber \\
=\frac{x^3}{5} [-10-20x-30x^2-12x^3+6x^4 + 9x^5 + 12x^6 + 15x^7+12x^8+9x^9+6x^10+3x^11) \nonumber \\
>0 \;\mbox{for} \; x>1 \mbox{~~~~~~~~~~~~~~~~~~~~~~~~~~~~~~~~~~~~~~~~~~~~~~~~~~~~~~~~~~~~~~~~~~~~~~~~~~~~~~~~~} \nonumber 
\end{eqnarray*}

\section*{Declaration of competing interest}
The authors declare that they have no known competing financial interests or personal relationships that could have appeared
to influence the work reported in this paper

\section*{acknowledgement}
One of the authors, KKG would like to thank the authority of Abacus Institute of Engineering and Management for all the facilities and encouragement. We all are grateful to the anonymous referee for the useful comments which have enhanced the quality of the paper.

\end{document}